\documentstyle[aps]{revtex}
\begin{document}
\newcommand{\bb}{\begin{eqnarray}}
\newcommand{\ee}{\end{eqnarray}}
\title{Is there still a strong CP problem?}
\author{H. Banerjee,} 
\author{D. Chatterjee}
\author{and}
\author{P. Mitra\thanks{mitra@tnp.saha.ernet.in}}
\address{Saha Institute of Nuclear Physics, 1/AF Bidhannagar,
Calcutta 700064}
\maketitle
\begin{abstract}The r\^{o}le 
of a chiral U(1) phase in the quark mass in QCD is analysed from first 
principles. In operator formulation, there is a 
parity symmetry and the phase can be removed by a change in 
the representation of the Dirac $\gamma$ matrices. Moreover,  
these properties are also realized in
a Pauli-Villars regularized version of the theory.
In the
functional integral scenario, attempts to remove the chiral phase by a chiral
transformation are thought to be obstructed 
by a nontrivial Jacobian arising from
the fermion measure and the chiral phase may therefore seem to
break parity.
But if one starts from the regularized action  
with the chiral phase also present in the regulator mass term, 
the Jacobian for a combined chiral rotation of quarks and regulators
is seen to be trivial and the phase
can be removed by a combined chiral
rotation. 
This amounts to a taming of the strong CP problem.
\end{abstract}

\section{Introduction}
The strong interactions do not violate parity in the manner of the weak
interactions: the gauge interactions in quantum chromodynamics involve vector
currents instead of chiral currents. Experiments too do not indicate
CP violation in the QCD sector. However, if one writes the Lagrangian
density as
\bb
{\cal L}=\bar\psi(i\gamma^\mu D_\mu-me^{i\theta'\gamma_5})\psi
-\frac{1}{4}~{\rm tr}~F_{\mu\nu}F^{\mu\nu}
-n_f{g^2\theta\over
32\pi^2}~{\rm tr}~F_{\mu\nu}\tilde F^{\mu\nu},
\label{L}\ee
it contains the topological term with the
QCD vacuum angle $\theta$, which violates CP and, if
nonzero, may lead to experimentally detectable CP violating effects.
Furthermore, the quark mass term has a chiral phase
$\theta'$, which comes from symmetry breaking in
the electroweak sector and may be large ($\approx 1$).
This phase is expected to violate CP by mixing a scalar and a pseudoscalar.
One then has to require that the CP violation due to $\theta'$
should be cancelled by that due to the $\theta$ term. 
In the popular functional integral scenario, the phase $\theta'$
is supposed to be convertible into a $\theta$-like term
through anomalous chiral rotations of the quarks, so that there is
an effective parameter $\bar\theta$ 
and according to popular methods of calculation the CP-violating
electric dipole moment of the neutron is thought to be 
about $10^{-16}\bar\theta$
e-cm, to be compared with the experimental upper bound of $10^{-26}$ e-cm.
This is interpreted to mean that $|\bar\theta|<10^{-10}$, 
requiring an unbelievable amount of fine-tuning between $\theta'$ and $\theta$.
This is referred to as the strong CP problem \cite{scp}.
Various modifications of QCD continue to be 
proposed to avoid this supposed need
for fine tuning. But none of them has been experimentally verified.

In such a situation it may be profitable to reanalyse the problem 
from first principles. 
There have been several attempts to prove that the $\theta$ parameter
does not really lead to any CP violation in QCD \cite{tokarev,sachs,lee}.  
As the nonperturbative dynamics of QCD with the topological term is difficult,
we find it useful to postpone the consideration of $\theta$ and 
concentrate on CP violation generated from the fermionic $\theta'$ term. 
We use a single flavour for simplicity. This means that $m$
in (\ref{L}) is a (real) number instead of a matrix. An extension can be 
made to more quark flavours without difficulty.

In first order, $\theta'$ is involved in an apparently pseudoscalar term, and
its matrix element has therefore been thought to be a measure of
the CP violation produced by the term. However, it is not so simple.
In theories like QCD with chirally invariant interactions, 
perturbation theory cannot
find any effect of such phases as they cancel at each vertex \cite{bcm}.
Thus the quark propagator can be written as
\bb
(\gamma^\mu p_\mu -me^{i\theta'\gamma_5})^{-1}
=e^{-i{\theta'\over 2}\gamma_5}
(\gamma^\mu p_\mu -m)^{-1} e^{-i{\theta'\over 2}\gamma_5},
\ee
and the chiral phases formally cancel at each gauge vertex because
\bb
e^{-i{\theta'\over 2}\gamma_5} \gamma^\mu e^{-i{\theta'\over 2}\gamma_5}=\gamma^\mu.
\ee
Moreover, in spite of the presence of the
chiral phase in the quark mass term, parity can be defined
so as to be conserved at the classical level.
This parity transformation, which leaves
the fermionic part of the Lagrangian $\int d^3x {\cal L}_\psi$ invariant
\cite{bcm}, involves the
usual parity operation for gauge fields, while the operation for fermions
includes a chiral rotation:
\bb
\bar\psi(x_0,\vec x)&\rightarrow &\bar\psi(x_0,-\vec x)
e^{i\theta'\gamma_5}\gamma^0\nonumber\\
\psi(x_0,\vec x)&\rightarrow &
\gamma^0e^{i\theta'\gamma_5}\psi(x_0,-\vec x).
\label{parity}\ee
The full mass term, with the chiral phase, is a scalar under this parity.

In spite of these direct and transparent arguments,  
the view that the chiral U(1) phase $\theta'$ in the quark
mass term gives rise to CP violating effects prevails in the literature. 
The raison d'etre
of this view is the perception that there is an anomaly in the axial U(1)
current:
\bb
\partial_\mu (\bar\psi\gamma_\mu\gamma_5\psi)
\stackrel{?}{\propto} {\rm tr} [F_{\mu\nu}\tilde{F}_{\mu\nu}] 
\ee
in euclidean metric. But it must be pointed out that
this anomaly does not translate $\theta'$ into a physically relevant
parameter unless supplemented by a nontrivial topology of the gauge field:
\bb
\int d^4x {g^2\over 16\pi^2}{\rm tr} [F_{\mu\nu}\tilde{F}_{\mu\nu}]=2\nu
\neq0. 
\ee
The fundamental point to be observed is that these two relations
are contradictory: the anomaly has the simple form only when $\nu=0$.
In the presence of a fermion mass, one must have
\bb
\int d^4x\langle
\partial_\mu (\bar\psi\gamma_\mu\gamma_5\psi)\rangle=0 \label{zero}
\ee
for consistency with a nontrivial topological charge because the
anomaly equation also contains a mass term, whose integral is 
\bb
\int d^4x [2m\langle\bar\psi\gamma_5\psi\rangle]=2(n_+-n_-)=2\nu,
\ee
by the index theorem. Here $n_{\pm}$ count the zero modes of the Dirac
operator with positive and negative chiralities. 
The theory becomes undefined unless the infrared divergences arising from
these zero modes are suitably regularized. In the infrared regularized
theory, the 
$m\to0$ limit of the LHS is nonvanishing, so that
the anomaly equation for massless fermions contains an extra piece
involving the zero modes \cite{hb} in the form 
$\sum_i\epsilon_i\phi^\dagger_{0i}(x) \phi_{0i}(x)$, 
where $\phi_{0i}$ is the $i$th zero mode eigenfunction
of the euclidean Dirac operator and $\epsilon_i$ its chirality. 
By virtue of (\ref{zero}) which continues to hold on infrared regularization, 
$\theta'$ becomes unphysical.

It has therefore to be examined carefully whether the abovementioned formal 
parity properties of the theory survive regularization and anomalies. 
We shall first review in the next section how the chiral
anomaly formally appears to lead to the breaking of parity
by $\theta'$. 
We shall thereafter introduce a (Pauli-Villars) regularization
and demonstrate that it is compatible with the parity defined in 
(\ref{parity}). In other words, $\theta'$ does {\it not}
cause any violation of parity in the regularized quantum theory. 
Then all that is needed for removing CP violation is to set $\theta=0$. 
This is natural in the technical sense \`{a} la 't Hooft, 
because it increases the CP symmetry of the
action and does not involve fine tuning between two quantities. 

\section{Anomaly and chiral rotation in unregularized theory}
   
The popular belief in the physicality of the $\theta'$ term 
and its equivalence with
a vacuum angle like $\theta$ arises mainly because of a nontrivial
Jacobian produced by a spacetime independent chiral
transformation in the euclidean functional integral. For a chiral transformation
\bb
\bar\psi\rightarrow\bar\psi e^{i\alpha(x)\gamma_5},\quad
\psi\rightarrow e^{i\alpha(x)\gamma_5}\psi,
\label{chi}\ee
where $\alpha$ may depend on $x$, the Jacobian reads
\bb
J=e^{i\int d^4x\alpha(x)X(x)},
\ee
where $X(x)\equiv\sum_n\phi^\dagger_n(x)\gamma_5\phi_n(x)$, the
functions $\phi_n$ being eigenfunctions of the Dirac operator in
euclidean metric.  A {\it regularized} calculation yields \cite{fujikawa}
\bb
X(x)={g^2\over 16\pi^2}{\rm tr} F_{\mu\nu}\tilde{F}_{\mu\nu}.
\ee
$X(x)$ is identified as the chiral anomaly \cite{fujikawa}
from the anomalous Ward identity
\bb
\langle\partial_\mu (\bar\psi(x)\gamma_\mu\gamma_5\psi(x))\rangle=
-2\langle\bar\psi m
\gamma_5\psi\rangle+X(x),
\label{ward}\ee
which follows from the rule that the functional integral over
$\psi,\bar\psi$ cannot change under a chiral rotation of integration variables.

If $\alpha(x)=-\theta'/2$, the chiral phase gets removed from the mass term,
but the Jacobian causes a parity violating vacuum 
angle term to be added to the action:
\bb
Z^{[\theta']}&\equiv&\int d\psi d\bar\psi e^{-\int d^4x \bar\psi(\gamma^\mu 
D_\mu-me^{i\theta'\gamma_5})\psi}\nonumber\\
&=&Z^{[0]}e^{-i\theta'{g^2\over 32\pi^2}
\int d^4x {\rm tr} F_{\mu\nu}\tilde{F}_{\mu\nu}}.
\label{z}\ee

Unlike the perturbative argument or the parity constructed earlier, this
functional integral argument takes the anomaly into account,
and hence it may appear to be more robust. 
However, a regularization is used here only in the evaluation of $X$, while
the remaining pieces of (\ref{ward})  are unregularized.  
We shall consider a fully regularized theory to resolve the apparent
discrepancy between the arguments of the previous section and the present one.
In the euclidean functional integral of the {\it regularized}
theory, it will turn out that 
the Jacobian corresponding to the physical fermions
can be cancelled by that corresponding to the regulators 
although both are separately nontrivial because 
of the existence of the chiral anomaly.

\section{Regularized QCD}
In the generalized Pauli-Villars regularization,
the Lagrangian density has to be augmented to include some extra species:
\bb
{\cal L}_{\psi,~ reg}^{[0]}=
\bar\psi(i\gamma^\mu D_\mu-m
)\psi+
\sum_j\sum_{k=1}^{|c_j|} \bar\chi_{jk}(i\gamma^\mu D_\mu-M_j
)\chi_{jk}.\label{PV}
\ee
Here the $\chi_{jk}$ are regulator spinor fields with 
fermionic or bosonic statistics, which determines 
the signs, positive or negative, of the integers $c_j$ \cite{faddeev};
the $c_j$-s have to satisfy relations
\bb
1+\sum_j c_j=0,\quad m^2+\sum_j c_jM_j^2=0 
\label{c}\ee
to cancel divergences.
The masses $M_j$ 
are to be taken to infinity at the end of calculations.

\subsection{Parity in regularized QCD in presence of chiral phase}
The easiest proof of the unphysicality of $\theta'$ is in the operator 
formulation of quantum chromodynamics.
Instead of an anomalous chiral transformation of the quark fields,
a change of $\gamma$-matrices can also be used to remove the chiral phase $\theta'$.
Using the relation
\bb
\bar\psi=\psi^\dagger\gamma^0,
\ee
one can write the spinor part of Eq. (\ref{L})  as
\bb
{\cal L}_\psi^{[\theta']}=\psi^\dagger(i\gamma^0\gamma^\mu D_\mu-
m\gamma^0e^{i\theta'\gamma_5})\psi,
\ee
where $D_\mu,m,\theta'$ carry no spinorial indices. It can also be rewritten as
\bb
{\cal L}_\psi^{[\theta']}=\psi^\dagger(i\widetilde{\gamma^0}\widetilde{\gamma^\mu}
D_\mu-m \widetilde{\gamma^0})\psi,
\ee
where
\bb
\widetilde{\gamma^\mu}\equiv e^{-i\theta'\gamma_5/2}{\gamma^\mu}
e^{i\theta'\gamma_5/2}.
\ee
It is easy to see that the new
matrices satisfy the Dirac algebra 
\bb
\widetilde{\gamma^\mu}\widetilde{\gamma^\nu}+
\widetilde{\gamma^\nu}\widetilde{\gamma^\mu}=2g^{\mu\nu}
\ee
and
also have the same hermiticity properties as their parent matrices.
Thus the chiral phase can be absorbed in a simple
redefinition of $\gamma$-matrices, and can have no physical effect.
Note that the parity transformation (\ref{parity}) is the usual parity
in terms of the $\widetilde{\gamma}$ representation. 

This argument goes through as long as there are no
additional chirally noninvariant 
interactions besides the one generating fermion mass.
It is also to be noted that this argument does not go through 
directly in euclidean field
theory, where $\bar\psi$ is taken to be independent of $\psi$. One 
can however modify the argument so that it holds in euclidean spacetime.
As it is Minkowski spacetime that one is finally interested in,
we shall not go into this complication here.

The argument skirts the issue of the chiral anomaly arising from
short distance singularities at the quantum level. We therefore need to
examine whether or not it is jeopardized in
the regularized theory.
The regularization (\ref{PV}) is standard, but in the presence of the 
chiral phase $\theta'$ 
in the physical fermion mass term, it is convenient to use the freedom to
provide the same chiral phase in the regulator mass terms as well:
\bb
{\cal L}_{\psi,~ reg}^{[\theta']}=
\bar\psi(i\gamma^\mu D_\mu-m
e^{i\theta'\gamma_5})\psi+
\sum_j\sum_{k=1}^{|c_j|} \bar\chi_{jk}(i\gamma^\mu D_\mu-M_j
e^{i\theta'\gamma_5})\chi_{jk}.\label{reg}
\ee
This is invariant under the parity transformation (\ref{parity}) 
extended to the regulator fields besides the physical fermion. Moreover, 
all the $\gamma_5$-s in (\ref{reg}) can be removed by changing 
over to the matrices
$\widetilde{\gamma^\mu}$ because the same chiral phase has been
chosen for all $j,k$: 
\bb
{\cal L}_{\psi,~ reg}^{[\theta']}=
\psi^\dagger\widetilde{\gamma^0}(i\widetilde{\gamma^\mu} D_\mu-m
)\psi+
\sum_j\sum_{k=1}^{|c_j|} \chi_{jk}^\dagger\widetilde{\gamma^0}
(i\widetilde{\gamma^\mu} D_\mu-M_j
)\chi_{jk}.
\ee
This is no more parity violating than (\ref{PV}).
Thus parity is conserved and $\theta'$ is unphysical in the regularized theory. 

If (\ref{reg}) were modified by changing the
chiral phases in the regulator sector, it would no longer respect the 
na\"{\i}ve extension 
of (\ref{parity}) to the regulators mentioned above. Such a choice
of phases would be equivalent to choosing nonzero phases for the 
regulators in the absence of a phase in the physical mass term.
However, it would still be possible to define a parity respected by
the regularized action by using different phases for the physical 
and regulator masses.

\subsection{Jacobian for chiral rotation in regularized functional integral}
The preceding arguments for the unphysicality of $\theta'$
suggest that something is amiss
in the anomaly-based argument leading to (\ref{z}). 
It is easy to guess what it may be: 
the functional integral in (\ref{z}) is not properly regularized.
The term $X(x)$, representing the anomaly, is calculated \cite{fujikawa}
through an {\it a posteriori} regularization. 
In this section, we start instead from the regularized Lagrangian density
(\ref{reg}).
The measure of integration now includes the Pauli-Villars fields: 
\bb 
d\mu=d\psi d\bar\psi\prod_{jk} d\chi_{jk} d\bar\chi_{jk}.
\ee
The fermionic functional integral is defined by
\bb
Z_{reg}^{[\theta']}\equiv\int d\mu e^{-\int d^4x 
{\cal L}_{\psi,~ reg}^{[\theta']}}.
\ee

One can apply a chiral transformation purely to the physical fermion fields 
and obtain formulas similar to (\ref{z}). 
However, this removes only the phase in the physical fermion fields 
in $Z_{reg}^{[\theta']}$, leaving behind some  
$\theta'$-dependence through the regularization. To remove both, one has to 
{\it extend} the chiral transformation (\ref{chi}) to the regulators. 
The Jacobian factors corresponding to the different $\chi_{jk},\bar\chi_{jk}$
come with powers $c_j/|c_j|$
by virtue of the fermionic or bosonic statistics, 
while $\psi,\bar\psi$ obey fermionic statistics. Altogether,
\bb
J_{reg}=e^{i(1+\sum_jc_j)\int d^4x \alpha(x)X(x)}=1,
\ee
because $1+\sum_j c_j=0$  \cite{fujikawa}.
Thus in the regularized framework the Jacobian 
for a {\it combined} chiral transformation on the physical fermion
fields and the regulators is trivial. 

Because of the trivial Jacobian associated with a combined
chiral rotation, we see that the fermionic functional integral
$Z_{reg}^{[\theta']}$
can be simplified by rotating $\theta'$ away
from both the physical fermion fields and the regulators:
\bb
Z_{reg}^{[\theta']}=Z_{reg}^{[0]},
\ee
{\it i.e.,} the regularized theories with and without $\theta'$ are equivalent.
In other words, the chiral phase $\theta'$ is completely unphysical
in the regularized theory.
The unphysicality of the chiral phase $\theta'$ proved earlier in operator
formulation thus stands vindicated in the framework of 
regularized functional integrals in the euclidean metric.

The lesson of the above exercise in the Pauli-Villars scheme is that a
regularized theory signals a trivial Jacobian and conversely,
a non-trivial Jacobian signals an unregularized theory.
What happens to the chiral anomaly in this regularized 
approach is an interesting question, which we now briefly discuss.

\subsection{Anomaly in regularized functional integral}
The regularized axial Ward identity is obtained from
the combined chiral transformation acting on physical fermions
and regulators: 
\bb
\langle\partial_\mu J_{\mu 5~reg}\rangle=
-2m\langle\bar\psi
\gamma_5\psi\rangle
-2\sum_{jk}M_j\langle\bar\chi_{jk}
\gamma_5\chi_{jk}\rangle,
\label{Ward}\ee
where
\bb
 J_{\mu 5~reg}=
\bar\psi(x) \gamma_\mu\gamma_5\psi(x)+
\sum_{jk}\bar\chi_{jk}(x) \gamma_\mu
\gamma_5\chi_{jk}(x).
\ee
Both sides of (\ref{Ward}) are regularized and well-defined.
One may wonder how the anomaly is manifested in this regularized approach
where the Jacobian for the combined chiral rotation is trivial
and the Ward identity does not look like an anomalous one.
In the regularized Ward identity (\ref{Ward}), 
divergent pieces in the physical mass term are cancelled exactly
by those in the regulator terms. As the regularization is removed \cite{brown},
\bb
-2\sum_{jk}\lim_{M_j\to\infty}M_j\langle\bar\chi_{jk}
\gamma_5\chi_{jk}\rangle_{reg}=
{g^2\over 16\pi^2}{\rm tr} F_{\mu\nu}\tilde{F}_{\mu\nu},
\ee
yielding the chiral anomaly and the regularized version of (\ref{ward}):
\bb
\langle\partial_\mu J_{\mu 5~reg}\rangle=
-2m\langle\bar\psi
\gamma_5\psi\rangle_{reg}+
{g^2\over 16\pi^2}{\rm tr} F_{\mu\nu}\tilde{F}_{\mu\nu},
\label{anomaly}\ee

\section{Discussion}
We have presented detailed arguments in this paper for the
unphysical nature of the phase $\theta'$.
One proof works in Minkowski spacetime and
involves a change of $\gamma$-matrices.  As is well known, no
representation of $\gamma$-matrices is more sacred than others, and as the 
phase can be removed by a mere change of representation, it cannot be 
physically observable. Another way of understanding this is to see 
that while the usual parity
operation involves $\gamma^0$, the presence of a chiral phase in the
mass term does not destroy the symmetry but merely
changes this matrix factor to $\widetilde{\gamma^0}$.
This new symmetry transformation involves a chiral transformation of the 
fermion fields, which alerts us to the possibility of a parity anomaly. 
We have used a generalized Pauli-Villars regularization 
to explore this possibility. It is consistent with 
the symmetry, showing that there is no parity anomaly.
An historically important approach involves a chiral transformation in which 
the chiral phase in the quark term is rotated away.
It is convenient for this purpose to make the same 
choice of chiral phase in the regulator sector as in the physical mass term.
In the regularized theory, the regulator fields 
are also chirally rotated and the chiral phase is fully rotated away:
the measure of the regularized euclidean functional integral does not 
change in this combined transformation, and no $F\tilde F$ term gets generated,
so that the theories with and without $\theta'$ are equivalent.
While the present paper uses an explicit regularization, the result
can alternatively be understood by considering the parity symmetry of the 
fermion functional measure \cite{measure}.

It is not difficult to understand why $\theta'$ was so long thought to
lead to CP violation. The formal functional integral formulation
chooses to ignore the regularization of the action and goes for a
regularization of only the change in the measure, which leads to a 
$\theta'$-dependent Jacobian modifying the $F\tilde F$ term.
A regularized action ensures that the functional integrals are well-defined and
yields a trivial Jacobian for the combined chiral transformation 
while the chiral anomaly is of course unchanged.

CP violation has been usually believed to get a 
contribution from the chiral phase
$\theta'$ of the quark mass term as well as the 
$\theta$ term in the gauge sector.
We have shown that $\theta'$ does not really have any such effect.
What can one say about the $\theta$ term?
This term changes sign under the standard parity transformation of gauge
fields, and there does not appear to be any redefinition of parity which
can restore the symmetry. This term is known to be an exact
divergence, and does not have any effect in perturbation theory. Its effect
has sometimes been estimated by conversion to a $\theta'$ term. Since such
a conversion is now seen to be impossible in a regularized theory, this method
has to be abandoned as invalid. There have been arguments in the literature 
against CP violation by the $\theta$ term, 
but one can still look for new ways of
estimating CP violating quantities 
on the basis of the $\theta$ term.

Meanwhile, the unphysicality of $\theta'$ already has a bearing on
the strong CP problem. 
It used to be thought that an unnaturally fine tuned cancellation
occurs between the $\theta'$ and $\theta$ terms,
and this was precisely the strong CP problem \cite{scp}.
The present work demonstrates that the $\theta'$ term produces no
CP violation. Thus $\theta'$ becomes irrelevant and there is no need
for fine tuning $\theta$ to balance its effects. 
Even without knowing how parity-violating $\theta$ is, one can 
avoid all CP violation by having $\theta=0$ in
the $\theta F\tilde F$ term. This is no doubt a special value, but
this choice increases the symmetry of the action, because
$\theta'$ does not break the symmetry. Consequently
it is a {\it natural} choice according to 't Hooft's criterion.
The situation is similar to
working with Lorentz-invariant theories, setting possible noncovariant 
terms to zero by hand, exploiting {\it naturalness}.
Regularization of QCD does not force $\theta$ or the CP violation
to vanish, but certainly makes it
{\it natural} to avoid CP violation by setting $\theta$ equal to zero.


\begin{thebibliography}{99}
\bibitem{scp} See, for example, 
J. E. Kim, Phys. Rep. {\bf 150} (1987) 1
\bibitem{tokarev} V. F. Tokarev, Mod. Phys. Letters {\bf A8} (1993) 531
\bibitem{sachs} R. G. Sachs, Phys. Rev. Letters {\bf 73} (1994) 377
\bibitem{lee} T. Lee, Int. J. Mod. Phys. {\bf A16} (2001) 4321   
\bibitem{bcm} H. Banerjee, D. Chatterjee and P. Mitra, 
``Is there a strong CP problem?'', 
SINP/TNP/90-5 (1990); 
Zeit. f\"{u}r Phys. {\bf C62} (1994) 511 
\bibitem{hb} H. Banerjee, 
Ind .J. Phys. (Spl.) {\bf 71} (1997) 333;
``Chiral Anomalies In Field Theories,'' 
in {\it Quantum Field Theory: A twentieth century
profile} ed. A. N. Mitra,
(Indian National Science Academy, New Delhi, 2000) [hep-th/9907162]
\bibitem{fujikawa} K. Fujikawa, Phys. Rev. {\bf D21} (1980) 2848 
\bibitem{faddeev} See, for example, L. Faddeev and A. Slavnov, {\it Gauge
Fields}, Benjamin-Cummings, Reading (1980).
\bibitem{brown} L. Brown, R. Carlitz and C. Lee,
Phys Rev. {\bf D16} (1977) 417 
\bibitem{measure} P. Mitra, 
``T invariance of Higgs interactions in the standard model,''
hep-ph/0303190, to appear in Proceedings of PASCOS'03
\end{thebibliography}
\end{document}